\documentclass[aps,onecolumn,showpacs,nofootinbib,showkeys]{revtex4-2}
\usepackage[margin=3.15cm]{geometry}
\usepackage{feynmp-auto}

\RequirePackage[T1]{fontenc}

\usepackage{epsfig,graphicx,amsmath,amssymb,bm}
\RequirePackage{mathptmx}      
\usepackage{mathrsfs}
\usepackage{amssymb}
\RequirePackage{color}
\usepackage{changes}
\usepackage{slashed}
\usepackage{multirow}
\usepackage{scalerel}
\usepackage{tikz-feynman}
\usepackage{textcomp}
\usepackage{subcaption}
\usepackage[english]{babel}
\usepackage{float}
\RequirePackage{hyperref}
\hypersetup{
    linktocpage,
    colorlinks,
    citecolor=blue,
    filecolor=black,
    linkcolor=blue,
    urlcolor=blue,
}
\usepackage{nccmath}

\begin{document}

\title{
Description of the process $e^+ e^- \to \pi^0\pi^0 \gamma$ in the extended NJL model
}


\author{M.K. Volkov$^{1}$}\email{volkov@theor.jinr.ru}
\author{A.A. Pivovarov$^{1}$}\email{tex$\_$k@mail.ru}
\author{K. Nurlan$^{1,2}$}\email{nurlan@theor.jinr.ru}

\affiliation{$^1$ Bogoliubov Laboratory of Theoretical Physics, JINR, 
                 141980 Dubna, Moscow region, Russia \\
                $^2$ The Institute of Nuclear Physics, Almaty, 050032, Kazakhstan
                }   


\begin{abstract}
The cross section of the process $e^+ e^- \to \pi^0\pi^0 \gamma$ is calculated in the extended NJL quark model. The vector channel with the intermediate $\rho$, $\omega$, $\rho'$ and $\omega'$ mesons is considered. The main contribution is given by the intermediate radially excited $\rho'$ meson. The role of the channel with the intermediate scalar meson is also discussed.


\end{abstract}


\maketitle

\section{\label{Intro}Introduction}
The study of meson production processes in electron-positron annihilation is important for testing the predictions of quantum chromodynamics (QCD) in the nonperturbative region and provides valuable data for the calculation of the anomalous magnetic moment of the muon. The process $e^{+}e^{-} \to \pi^{0}\pi^{0}\gamma$ is of particular interest for a few reasons. Firstly, the final state $\pi^{0}\pi^{0}\gamma$ is an important source of information about radiative decays of light vector mesons such as $(\rho, \omega, \phi) \to \pi^0\pi^0\gamma$ and their radial excitations~\cite{ParticleDataGroup:2024cfk}. Secondly, unlike the process $e^{+}e^{-} \to \pi^{0}\pi^{0}$ which is forbidden in the leading order, the channel with an isolated photon allows one to study the dynamics of strong interactions at the vertex $\gamma^{*} \to \pi^{0}\pi^{0}\gamma$, which is dominated by the contributions of intermediate vector resonances. Experimental measurements of this process were performed by the SND and CMD-2 collaborations at the VEPP-2000 collider in the energy range 1.05 - 2 GeV~\cite{Achasov:2000wy,CMD-2:2003bgh,Achasov:2013btb,Achasov:2016zvn}. 

The Nambu--Jona-Lasinio (NJL) model ~\cite{Ebert:1982pk,Volkov:1984kq,Volkov:1986zb,Ebert:1985kz,Vogl:1991qt,Klevansky:1992qe,Ebert:1994mf} is an effective tool for describing light meson interactions at low energies. In the framework of this model, one can describe the properties of ground-state mesons. The radially excited meson states can be described in the framework of its extended version~\cite{Volkov:1996br,Volkov:1996fk,Volkov:2005kw,Volkov:2017arr}. An important advantage of the NJL model is that calculations of the $V\pi\gamma$ or $V'\pi\gamma$ vertices do not require additional arbitrary parameters. They can be calculated by considering convergent quark loops of anomalous type~\cite{Volkov:2005kw,Volkov:2017arr}. This allows one to made predictions for the cross sections of the processes $e^{+}e^{-} \to \pi(\pi')\gamma$~\cite{Arbuzov:2011fv}, $e^{+}e^{-} \to [\eta,\eta',\eta(1295),\eta(1475)]\gamma$~\cite{Ahmadov:2013ksa} and others. In particular, the NJL model reproduces the result of the VMD model when summing channels with an isolated photon and the $\rho$ meson exchange.

In the present work, we give a theoretical description of the process \(e^{+}e^{-} \to \pi^{0}\pi^{0}\gamma\) in the framework of the extended NJL model. The aim of the work is to provide a theoretical prediction of the cross section of this process at energies up to 2 GeV taking into account the main intermediate states: the contact term via a virtual photon, vector mesons $\rho$, $\omega$, and their first radially excited states $\rho(1450)$ and $\omega(1420)$. Unlike the phenomenological model VMD where the vertex parameters and coupling constants are taken from experimental data, in our approach all the vertices (for example $g_{\rho \pi \gamma}$, $g_{\omega \pi \gamma}$, $g_{\rho' \pi \gamma}$, $g_{\omega' \pi \gamma}$) are calculated with coupling constants fixed at the level of the free Lagrangian. We will also discuss the role of different contributions to the cross section of the process and compare our results with the experimental data.
 
\section{Lagrangian of the NJL model} 
\label{sect:NJL}
The description of radially excited states in the exctended NJL model is carried out by using a form factor of the polynomial type \cite{Volkov:1996br,Volkov:1996fk}. In order to obtain physical states, the Lagrangian is diagonalized. As a result, the quark-meson Lagrangian of the extended NJL model containing the vertices necessary for the considered process takes the following form~\cite{Volkov:2005kw,Volkov:2017arr}:
\begin{eqnarray}
	\label{Lagrangian}
		\Delta L_{int} & = &
		\bar{q} \biggl[ i \gamma^{5} \sum_{j = \pm} \lambda_{j}^{\pi} \left(A_{\pi}{\pi}^{j} + B_{\pi}{\pi'}^{j}\right) +
		+\frac{1}{2} \gamma^{\mu} \sum_{j = \pm} \lambda_{j}^{\rho} \left(A_{\rho}\rho^{j}_{\mu} + B_{\rho}{\rho'}^{j}_{\mu} \right) \nonumber \\ 
		&& 
		+\frac{1}{2} \gamma^{\mu} \sum_{j = \pm} \lambda_{j}^{\omega} \left(A_{\omega}\omega^{j}_{\mu} + B_{\omega}{\omega'}^{j}_{\mu} \right)
		\biggl]q,
\end{eqnarray}
where $q$ is a doublet of $u$ and $d$ quarks, and $\lambda$ are the linear combinations of the Gell-Mann matrices. The factors $A_M$ and $B_M$ contain the mixing angels:
\begin{eqnarray}
\label{verteces1}
	A_{M} = \frac{1}{\sin(2\theta_{M}^{0})}\left[g_{M}\sin(\theta_{M} + \theta_{M}^{0}) +
	g'_{M}f(k_{\perp}^{2})\sin(\theta_{M} - \theta_{M}^{0})\right], \nonumber\\
	B_{M} = \frac{-1}{\sin(2\theta_{M}^{0})}\left[g_{M}\cos(\theta_{M} + \theta_{M}^{0}) +
	g'_{M}f(k_{\perp}^{2})\cos(\theta_{M} - \theta_{M}^{0})\right],
\end{eqnarray}
where $M$ denotes the appropriate meson. The values of the mixing angles are given in Table~\ref{tab_mixing}.
\begin{table}[h!]
\caption{The values of the mixing angles of the mesons~\cite{Volkov:2017arr}.}
\begin{center}
\begin{tabular}{cccc}
\hline
   & $\pi$ & $\rho$ & $\omega$ \\
\hline
$\theta_M$	& $59.48^{\circ}$	&  $81.80^{\circ}$  & $81.80^{\circ}$ \\
$\theta^0_M$	& $59.12^{\circ}$	& $61.50^{\circ}$  & $61.50^{\circ}$ \\
\hline
\end{tabular}
\end{center}
\label{tab_mixing}
\end{table}

The function $f(k_{\perp}^2) = 1 + d k_{\perp}^2$ is the form factor introduced to describe the first radially excited meson states. $d$ is the slope parameter determined from the requirement that the introduction of the excited states does not change the quark condensate. 

The coupling constants are
\begin{eqnarray}
\label{Couplings}
 g_{\pi} = \left(\frac{4}{Z_{\pi}}I_{2}\right)^{-1/2}, &\quad&
\, g'_{\pi} =  \left(4 I_{2}^{f^{2}}\right)^{-1/2}, \nonumber\\
g_{\rho} = g_{\omega} = \left(\frac{2}{3}I_{2}\right)^{-1/2}, &\quad&
\, g'_{\rho} = g'_{\omega} =\left(\frac{2}{3}I_{2}^{f^{2}}\right)^{-1/2}
\end{eqnarray}
where $Z_\pi = \left(1 - 6 \frac{m^2}{M_{a_1}^2}\right)^{-1}$ is the additional renormalization constant appearing from taking into account the $\pi-a_1$ transitions.

The integrals appearing in the renormalization of the Lagrangian are
\begin{eqnarray}
	I_{n}^{f^{m}} =
	-i\frac{N_{c}}{(2\pi)^{4}}\int\frac{f^{m}(k^2_{\perp})}{(m^{2} - k^2)^{n}}\Theta(\Lambda_{3}^{2} - k^2_{\perp})
	\mathrm{d}^{4}k.
\end{eqnarray}
where $m = 270$~MeV is the mass of the $u$ quark, $\Lambda_3=1030$~MeV is the cut-off parameter~\cite{Volkov:2017arr}.
   
\section{The amplitude and cross section of the process $e^+e^- \to \pi^0 \pi^0 \gamma$} 
	
    \begin{figure*}[t]
 \centering
   \centering
   \begin{tikzpicture}
    \begin{feynman}
      \vertex (a) {\(e^+\)};
      \vertex [dot, below right=1.8cm of a] (b){};
      \vertex [below left=1.8cm of b] (c) {\(e^-\)};
      \vertex [blob, right=1.4cm of b] (d) {};
      \vertex [above right=1.8cm of d] (e) {\(\pi^0\)};
      \vertex [below right=1.8cm of d] (h) {\(\gamma\)};
      \vertex [right=1.5cm of d] (f) {\(\pi^0\)};
      \diagram* {
         (a) -- [fermion] (b),
         (b) -- [fermion] (c),
         (b) -- [boson, edge label'=\({\gamma^*}\)] (d),    
         (d) -- [double] (e),  
         (d) -- [double] (h),
         (d) -- [double] (f),
      };
     \end{feynman}
    \end{tikzpicture}
   \caption{The Feynman diagram with an intermediate photon. The dashed circle represents the sum of sub-diagrams given in Fig. \ref{diagram3}.}
 \label{diagram1}
\end{figure*}
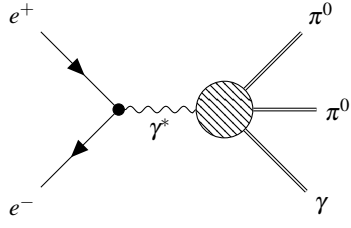%
\begin{figure*}[t]
 \centering
   \centering
   \begin{tikzpicture}
    \begin{feynman}
      \vertex (a) {\(e^+\)};
      \vertex [dot, below right=1.8cm of a] (b){};
      \vertex [below left=1.8cm of b] (c) {\(e^-\)};
      \vertex [dot, right=1.0cm of b] (d) {};
      \vertex [dot, right=0.8cm of d] (l) {};
      \vertex [blob, right=1.4cm of l] (g) {};
      \vertex [above right=1.8cm of g] (e) {\(\pi^0\)};
      \vertex [below right=1.8cm of g] (h) {\(\gamma\)};
      \vertex [right=1.5cm of g] (f) {\(\pi^0\)};
      \diagram* {
         (a) -- [fermion] (b),
         (b) -- [fermion] (c),
         (b) -- [boson, edge label'=\({\gamma^*}\)] (d),
         (d) -- [fermion, inner sep=1pt, half left] (l),
         (l) -- [fermion, inner sep=1pt, half left] (d),
         (l) -- [double, edge label'=\({V, V'} \)] (g),
         (g) -- [double] (e),  
         (g) -- [double] (h),
         (g) -- [double] (f),
      };
     \end{feynman}
    \end{tikzpicture}
   \caption{The Feynman diagram with intermediate vector mesons $V=\rho, \omega$ and $V'=\rho', \omega'$. The dashed circle represents the sum of sub-diagrams given in Fig. \ref{diagram3}.}
 \label{diagram2}
\end{figure*}
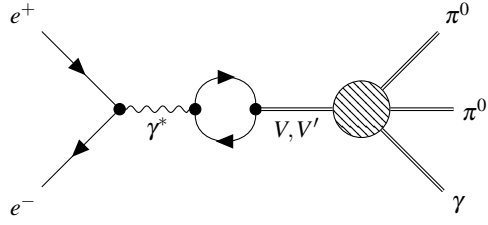%

    \begin{figure*}[t]
 \centering
  \begin{subfigure}{0.5\textwidth}
   \centering
   \begin{tikzpicture}
     \begin{feynman}
      \vertex [dot] (d) {};      
      \vertex [dot, above right=1.4cm of d] (e) {};
      \vertex [dot, below right=1.4cm of d] (h) {};
      \vertex [dot, right=1.2cm of e] (f) {};
      \vertex [dot, above right=1.2cm of f] (n) {};  
      \vertex [dot, below right=1.2cm of f] (m) {};   
      \vertex [right=1.2cm of n] (l) {\(\ \pi^0 \)}; 
      \vertex [right=1.2cm of m] (s) {\(\gamma \)};  
      \vertex [right=1.4cm of h] (k) {\(\ \pi^0 \)}; 
      \diagram* {         
         (d) -- [fermion] (e),  
         (e) -- [fermion] (h),
         (d) -- [anti fermion] (h),
         (e) -- [double, edge label'=\({\rho, \omega} \)] (f),
         (f) -- [fermion] (n),
         (n) -- [fermion] (m),
         (f) -- [anti fermion] (m), 
         (h) -- [double] (k),
         (n) -- [double] (l),
	 (m) -- [double] (s),
      };
     \end{feynman}
    \end{tikzpicture}
  \end{subfigure}%
 \centering
 \begin{subfigure}{0.5\textwidth}
  \centering
   \begin{tikzpicture}
     \begin{feynman}
      \vertex [dot] (d) {};      
      \vertex [dot, above right=1.4cm of d] (e) {};
      \vertex [dot, below right=1.4cm of d] (h) {};
      \vertex [dot, right=1.2cm of e] (f) {};
      \vertex [dot, above right=1.2cm of f] (n) {};  
      \vertex [dot, below right=1.2cm of f] (m) {};   
      \vertex [right=1.2cm of n] (l) {\(\ \pi^0 \)}; 
      \vertex [right=1.2cm of m] (s) {\( \gamma \)};  
      \vertex [right=1.4cm of h] (k) {\(\ \pi^0 \)}; 
      \diagram* {         
         (d) -- [fermion] (e),  
         (e) -- [fermion] (h),
         (d) -- [anti fermion] (h),
         (e) -- [double, edge label'=\({\rho', \omega'} \)] (f),
         (f) -- [fermion] (n),
         (n) -- [fermion] (m),
         (f) -- [anti fermion] (m), 
         (h) -- [double] (k),
         (n) -- [double] (l),
	 (m) -- [double] (s),
      };
     \end{feynman}
    \end{tikzpicture}
  \end{subfigure}%
 \caption{The vertex $V(V')\pi^0\pi^0\gamma$ with two anomalous triangular quark loops connected by virtual vector mesons in the ground and excited states $\rho$, $\rho'$, $\omega$ and $\omega'$.
 }
 \label{diagram3}
\end{figure*}
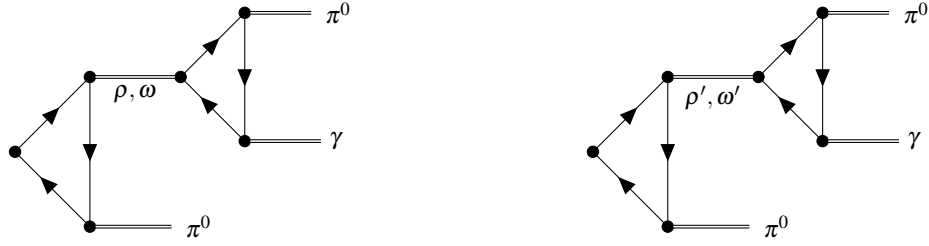%

    The process $e^+e^- \to \pi^0 \pi^0 \gamma$ is described by a contact diagram with a photon transition to the final states through anomalous vertices $\gamma^*\to\omega^{(\prime)}\pi\to\gamma\pi\pi$ and by a diagram with intermediate vector mesons in the ground and first-radially excited states. The corresponding diagrams are presented in Figs. \ref{diagram1} and \ref{diagram2}. The amplitude of the process is written as the sum of all contributions

	\begin{eqnarray}
	\mathcal{M}(e^+e^- \to \gamma\pi\pi) = \frac{128 {(\pi \alpha_{em})}^{3/2} m^2_u}{s} l_{\mu} \biggl[ {B_{(\rho + \rho')}}_{\mu\nu}  
    \left[I^{\omega\pi}_{30} \, BW_{\omega}(q^2_{\pi\gamma}) + 
    I^{\omega'\pi}_{30} \, BW_{\omega'}(q^2_{\pi\gamma})\right] +
    \\ \nonumber
    {B_{(\omega + \omega')}}_{\mu\nu}  
    \left[I^{\rho\pi}_{30} \, BW_{\rho}(q^2_{\pi\gamma}) + 
    I^{\rho'\pi}_{30} \, BW_{\rho'}(q^2_{\pi\gamma})\right]
    \biggl]
    \\ \nonumber
    \epsilon_{\nu\alpha\beta\lambda}\epsilon_{\delta\alpha\tau\sigma}
    p_{3\beta} {(p_2 +p_4)}_\lambda p_{2\tau} p_{4\sigma}
    e_\nu(\gamma) + (p_2\leftrightarrow p_3),
	\end{eqnarray}
	where $q^2 = (p(e^{-}) + p(e^{+}))^2$, $l^{\mu} = \bar{e}\gamma^{\mu}e$ is the lepton current. The momenta of the mesons in the final state and photons are defined as $p_{\pi^0}=p_2$, $p_{\pi^0}=p_3$, and $p_{\gamma}=p_4$. The terms in square brackets in the amplitude describe the contributions from the contact diagram and diagrams with intermediate vector mesons. The meson masses and widths are taken from PDG \cite{ParticleDataGroup:2024cfk}.

    The sum of the contributions of the vector $\rho$ and $\rho'$ mesons with the corresponding part of the contact diagram is
	\begin{eqnarray}
	&& B_{(\rho + \rho')\mu\nu} = g_{\mu\nu}I^{\omega^{(\prime)}\pi}_{30} + \frac{C_{\rho}}{g_{\rho}}\frac{g_{\mu\nu}q^2 - q_{\mu}q_{\nu}}{M^{2}_{\rho} - q^2 - i\sqrt{q^2}\Gamma_{\rho}} I^{\rho\omega^{(\prime)}\pi}_{30}  \nonumber + \frac{C_{\rho'}}{g_{\rho}}\frac{g_{\mu\nu}q^2 - q_{\mu}q_{\nu}}{M^{2}_{\rho'} - q^2 - i\sqrt{q^2} \Gamma_{\rho'}} I^{\rho' \omega^{(\prime)}\pi}_{30}.
	\end{eqnarray}

    The contribution of the channels with intermediate $\omega$ and $\omega'$ mesons has the form
	\begin{eqnarray}
	&& B_{(\omega + \omega')\mu\nu} = g_{\mu\nu} \frac{I^{\rho^{(\prime)}\pi}_{30}}{3} + \frac{C_{\omega}}{3g_{\rho}}\frac{g_{\mu\nu}q^2 - q_{\mu}q_{\nu}}{M^{2}_{\omega} - q^2 - i\sqrt{q^2}\Gamma_{\omega}} I^{\omega\rho^{(\prime)}\pi}_{30}  \nonumber + \frac{C_{\omega'}}{3g_{\rho}}\frac{g_{\mu\nu}q^2 - q_{\mu}q_{\nu}}{M^{2}_{\omega'} - q^2 - i\sqrt{q^2} \Gamma_{\omega'}} I^{\omega' \rho^{(\prime)}\pi}_{30},
	\end{eqnarray}
    where the constants $C_\rho=3C_\omega$ and $C_{\rho'}=3C_{\omega'}$ appear in the quark loops of the $\gamma \to \rho(\rho')$ transition
	\begin{eqnarray}
    \label{gammatransitions}
	C_{\rho} = A^0_\rho \biggl[\sin(\theta_{\rho} + \theta_{\rho}^{0})	+ R_{\rho}\sin(\theta_{\rho} - \theta_{\rho}^{0})\biggl], \nonumber\\
	C_{\rho'} = -A^0_\rho \biggl[\cos(\theta_{\rho} + \theta_{\rho}^{0}) + R_{\rho}\cos(\theta_{\rho} - \theta_{\rho}^{0})\biggl],
	\end{eqnarray}
    where $R_\rho \approx 0.55$ \cite{Volkov:2017arr}.

    For the intermediate $\rho'$ meson, the energy-dependent width is used \cite{CMD-2:2003bgh}
    \begin{eqnarray}
    \Gamma_{\rho'}(q^2) = \Gamma_{\rho'} \biggl[
    \text{Br}(\rho'\to\omega\pi) {\left( \frac{p_\omega(q^2)}{p_\omega(M^2_{\rho'})} \right)}^3 +
    \left[1 - \text{Br}(\rho'\to\omega\pi)\right] \frac{M^2_{\rho'}}{q^2}
    {\left( \frac{p_\pi(q^2)}{p_\pi(M^2_{\rho'})} \right)}^3
    \biggl], 
    \end{eqnarray}
    where we use $\text{Br}(\rho'\to\omega\pi)=31.2\,\%$ \cite{Volkov:2023hju}.

	The integrals arising in quark loops with meson vertices defined in (\ref{Lagrangian}) have the following form:
	\begin{eqnarray}
	\label{DiffIntegral}
		I_{n_{1} n_{2}}^{M, \dots, M^{'}, \dots} & = &
		-i\frac{N_{c}}{(2\pi)^{4}}\int\frac{A_{M} \dots B_{M} \dots}{(k^2 - m_{u}^{2})^{n_{1}}(k^2 - m_{s}^{2})^{n_{2}}} \Theta(\Lambda^{2} - k_{\perp}^2) \mathrm{d}^{4}k,
	\end{eqnarray}
	where $A_M$ and $B_M$ are defined in (\ref{verteces1}).

\begin{figure}[h]
    \center{\includegraphics[scale = 0.8]{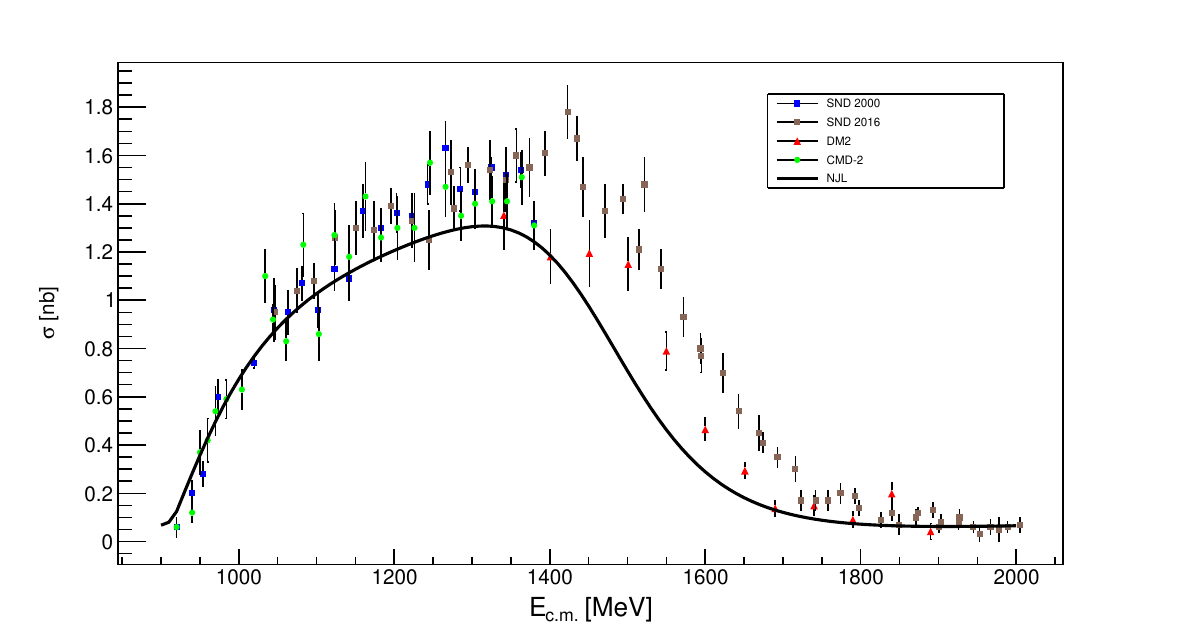}}
    \caption{
    Comparison of the $e^+e^- \to \gamma\pi^0\pi^0$ process total cross section calculated in the NJL model with experimental data of the SND \cite{Achasov:2000wy,Achasov:2016zvn}, CMD-2 \cite{CMD-2:2003bgh} and DM2 \cite{DM2:1990npw} collaborations.
    }
    \label{crosssection}
    \end{figure}

    The dependence of the cross section of the processes under consideration on the energy of colliding leptons obtained on the basis of the above amplitudes and their comparison with experimental data is presented in Fig.~\ref{crosssection}.
    
\section{Conclusion and discussion} 
In the present work, within the extended $SU(2)\times SU(2)$ NJL model, the cross-section of the process $e^{+}e^{-} \to \pi^{0}\pi^{0}\gamma$ is calculated in the center of mass energy range up to 2~GeV. The contributions of the contact channel and channels with intermediate vector mesons $\rho$, $\omega$, $\rho'(1450)$ and $\omega(1420)$ are taken into account. The model correctly reproduces the position of the main resonance associated with the contribution of the intermediate $\rho'(1450)$ meson, and also qualitatively describes the height of the resonance and the behavior of the cross section in the threshold region.

An important feature of the model is its ability to describe the cross section of a process involving vector meson channels in both the ground state and the first radially excited state without using any additional arbitrary parameters. The number of main parameters is limited and fixed at the stage of constructing the model's free Lagrangian. This makes it possible to predict the process cross sections.
The numerical calculations show that the exchange of a photon and a $\rho$-meson is important in the threshold region, while the contribution of $\rho^{\prime}$ with the further transition $\rho' \to \omega\pi$ dominates in the mass region $M_{\rho^{\prime}}=1.465$ GeV and mainly determines the cross section of the process. Taking into account the channel with the second excited state $\rho(1700)$ could possibly improve agreement with experimental data in the region above the $\rho(1450)$ resonance. At the same time, the analysis of the SND collaboration data showed an insignificant contribution of $\rho(1700)$ to the cross section of the $e^{+}e^{-} \to \pi^{0}\pi^{0}\gamma$ process in the energy region up to 2 GeV \cite{Achasov:2013btb,Achasov:2016zvn}. On the other hand, the NJL model describes mesons in the ground state and only in the first radially excited state. It should be noted that the model quite successfully described a number of $e^+e^-$ annihilation  processes into mesons and hadronic $\tau$ decays \cite{Volkov:2017arr}.

As noted in Introduction, the process $e^+e^- \to \pi^0\pi^0\gamma$ is important for understanding the radiative decays of $\rho \to \pi^0\pi^0\gamma$, $\omega \to \pi^0\pi^0\gamma$ and decays of excited states. Theoretical calculations of radiative decay widths in various theoretical approaches indicate an important role of the scalar channel $\rho \to f_0 \gamma\to\pi^0\pi^0\gamma$ in addition to the vector channel $\rho \to \omega\pi \to \pi^0\pi^0\gamma$ \cite{Oh:2003zz,Palomar:2001vg,Escribano:2006mb,Radzhabov:2007wk}. An experimental search for the cross section of the $e^+e^- \to f_0 \gamma$ processes and a study of its contribution to the $e^+e^- \to \pi^0\pi^0\gamma$ process was carried out in \cite{Achasov:2011zza}.
In the energy range 1.30 - 1.38 GeV, an upper limit on the cross section of the process $\sigma(e^+e^- \to f_0 \gamma) < 30$ pb at the level of 90\% CL was obtained. The contributions to the total cross section of channels with scalar and tensor mesons $e^+e^- \to f \gamma \to \pi^0\pi^0\gamma$ (where $f=f_0(500), f_0(980), f_0(1350), f_2(1270),$) were estimated as $0.5-3.2$\% of the process cross section $e^+e^- \to \omega \pi \to \pi^0\pi^0\gamma$ in this energy range. In that work, the limits on the product of relative decay probabilities $Br(\rho'\to e^+e^-) Br(\rho' \to f_0(500)\gamma) < 4.0 \times 10^{-9}$ were also obtained.
Using the $\rho'\to\gamma$ transition described by formula (\ref{gammatransitions}), one can obtain a theoretical estimate of the decay $\Gamma(\rho'\to e^+e^-) = 4\pi \alpha_{em} C_{\rho'}M_{\rho'} / 3 g^2_\rho = 1.01$ keV. Then the limit on the branching ratio of the decay with scalar meson is $Br(\rho'\to f_0(500)\gamma) < 1.6 \times 10^{-3}$. It should be noted that the partial width of the anomalous decay satisfies $Br(\rho'\to\omega\pi)=31.2\%$ \cite{Volkov:2023hju}. All this indicates a relatively small contribution from the channel with intermediate scalar mesons with the vertices $\rho f_0\gamma$, $\rho' f_0\gamma$ in the process $e^+e^- \to \pi^0\pi^0\gamma$.

\subsection*{Acknowledgments}
    The authors are grateful to Prof. A.~B.~Arbuzov for useful discussions.


\end{document}